\documentclass[prl,preprintnumbers,amsmath,amssymb,twocolumn]{revtex4}
\usepackage{graphicx}
\usepackage{bm}

\begin{document}
\title{Hitting probability for anomalous diffusion processes}
\author{Satya N.~Majumdar}
\affiliation{CNRS - Universit\'e Paris-Sud, LPTMS, UMR8626 - B\^at.~100, 91405 Orsay Cedex, France}
\author{Alberto Rosso}
\affiliation{CNRS - Universit\'e Paris-Sud, LPTMS, UMR8626 - B\^at.~100, 91405 Orsay Cedex, France}
\author{Andrea Zoia}
\email{andrea.zoia@cea.fr}
\affiliation{CEA/Saclay, DEN/DM2S/SERMA/LTSD, B\^at.~454, 91191 Gif-sur-Yvette Cedex, France}

\begin{abstract} We present the universal features of the hitting probability $Q(x,L)$, the probability that a generic stochastic process starting at $x$ and evolving in a box $[0,L]$ hits the upper boundary $L$ before hitting the lower boundary at $0$. For a generic self-affine process (describing, for instance, the polymer translocation through a nanopore) we show that $Q(x,L)=Q(x/L)$ and the scaling function $Q(z)\sim z^\phi$ as $z\to 0$ with $\phi=\theta/H$ where $H$ and $\theta$ are respectively the Hurst exponent and the persistence exponent of the process. This result is verified in several exact calculations including when the process represents the position of a particle diffusing in a disordered potential. We also provide numerical supports for our analytical results.

\end{abstract}

\maketitle

\begin{figure}
\centerline{
\includegraphics[width=.5\columnwidth]{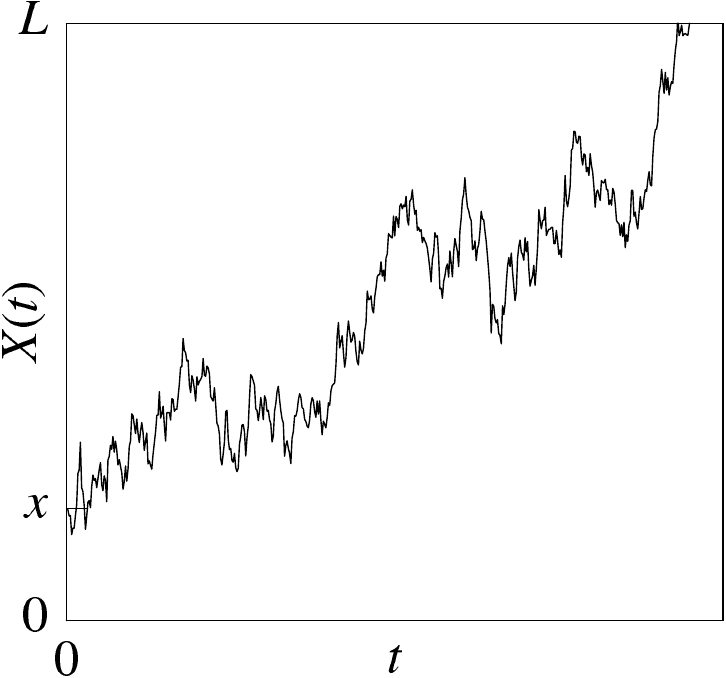}
\includegraphics[width=.5\columnwidth]{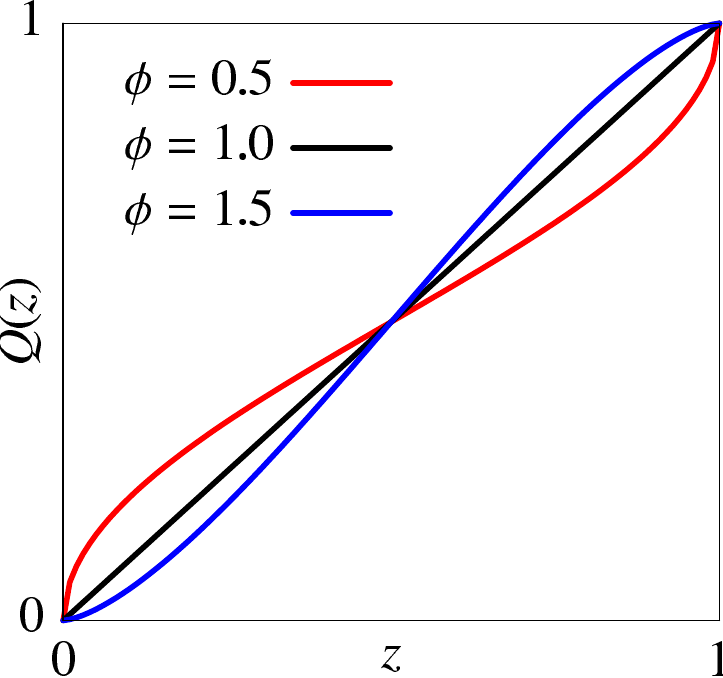}}
\caption{Left. The evolution of a stochastic process initiated at $X(0)=x$ and terminated upon exiting from the box of size $L$. Right. The function $Q(z)$ as given by Eq.~\eqref{incomplete_beta} for different values of the exponent $\phi$.}
\label{fig:fig1}
\end{figure}

The transfer of DNA, RNA and proteins through cell membranes is key to understanding several biological processes~\cite{ref1}. The transport of polymer molecules across nanopores is also relevant in many chemical and industrial applications~\cite{ref3}. A fundamental question concerns whether a polymer, once penetrated into the pore, will eventually complete its transit. The answer is naturally formulated in terms of the translocation coordinate $X(t)$, namely the  length of the translocated portion of the polymer at time $t$~\cite{ref8,joanny,kardar3,panja}. In absence of driving forces, the polymer dynamics is governed by thermal fluctuations. In this case, the traslocation coordinate can be expressed as a stochastic process $X(t)$ that evolves in a box of size $L$ ($L$ being the polymer length), starting from some initial value $X(0)=x$, $0<x <L$, and terminated upon touching either boundary for the first time (Fig.~\ref{fig:fig1} left). It has been shown that excluded volume effects hinder the polymer dynamics, so that the process $X(t)$ actually undergoes subdiffusion~\cite{ZRM,kardar3}. We define the hitting probability $Q(x,L)$ as the probability of exiting the domain through the boundary at $L$, which corresponds to the polymer completing the translocation.

More generally, the hitting probability $Q(x,L)$ of a particle undergoing anomalous (i.e., non-Brownian) diffusion is key to understanding a variety of phenomena, such as the classical gambler's ruin problem in finance and risk management~\cite{Feller,redner}, the transport of charge carriers in conductors with disordered impurities~\cite{bouchaud} and the breakthrough of chemical species in heterogeneous porous media for contaminated sites remediation~\cite{rev_geo}, only to name a few. For ordinary Brownian diffusion, the hitting probability $Q(x,L)=x/L$ is easy to compute~\cite{Feller,redner}. The goal of this Letter is to study $Q(x,L)$ for generic self-affine stochastic processes, thus going beyond the Brownian world.

Since the only length scale in the problem is $L$, evidently $Q(x,L)$ is a function of only the scaled variable $x/L$:$Q(x,L)=Q(x/L=z)$. For a Brownian motion, $Q(z)=z$, a simple linear function. For a generic stochastic process $X(t)$, $Q(z)$ is non trivial (see for example Fig.~\ref{fig:fig1} right). The central aim of this Letter is to determine the universal features associated with $Q(z)$ in the two following cases: symmetric self-affine processes, characterized by a power-law scaling $X(t)\sim t^H$, with Hurst exponent $H>0$; and a single particle diffusing in a disordered potential $V(X)$. The translocation process belongs to the former, whereas transport in quenched disorder to the latter.

It is useful to summarize our main results which are
threefold: $i)$ For self-affine processes, we show that 
generically $Q(z) \sim z^\phi$ for small $z$, where $\phi=\theta/H$, and 
$\theta$ is 
the so-called persistence exponent~\cite{pers_review} of the same process 
in a semi-infinite geometry. $ii)$ 
For a particle diffusing 
in a disordered potential $V(X)$, we provide an 
exact formula for $Q(x,L)$ valid for 
arbitrary $V(X)$ which incidentally also allows us to compute the 
persistence 
exponent of a 
particle diffusing 
in a self-affine disordered potential. $iii)$ The function $Q(z)$ is explicitly 
known for some anomalous diffusion processes. Amazingly, we find that these 
apparently different-looking formulae can be cast in the same 
super-universal form, when 
expressed in terms of the exponent $\phi$. This naturally raises the question: how generic is this super-universality? We provide numerical evidences that indeed in some cases the super-universality is violated and we discuss its limit of validity.

\begin{figure}
\centerline{
\includegraphics[width=.5\columnwidth]{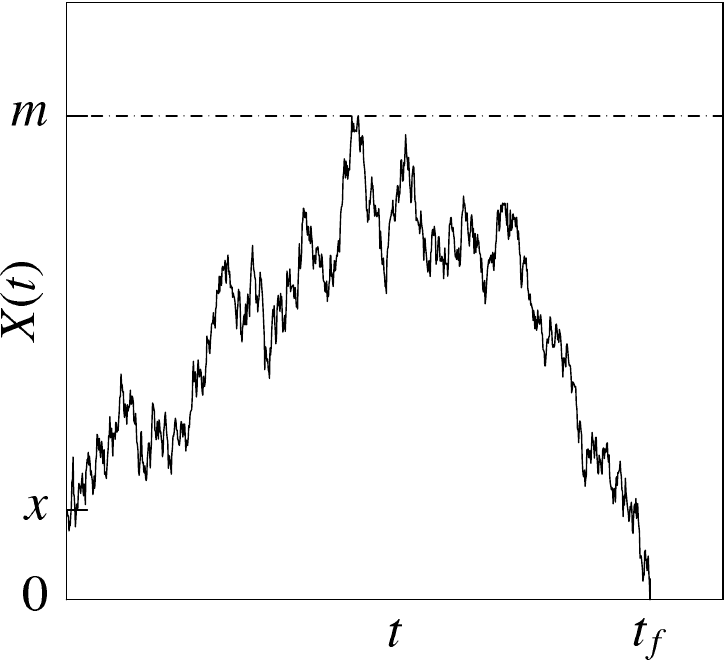}
\includegraphics[width=.5\columnwidth]{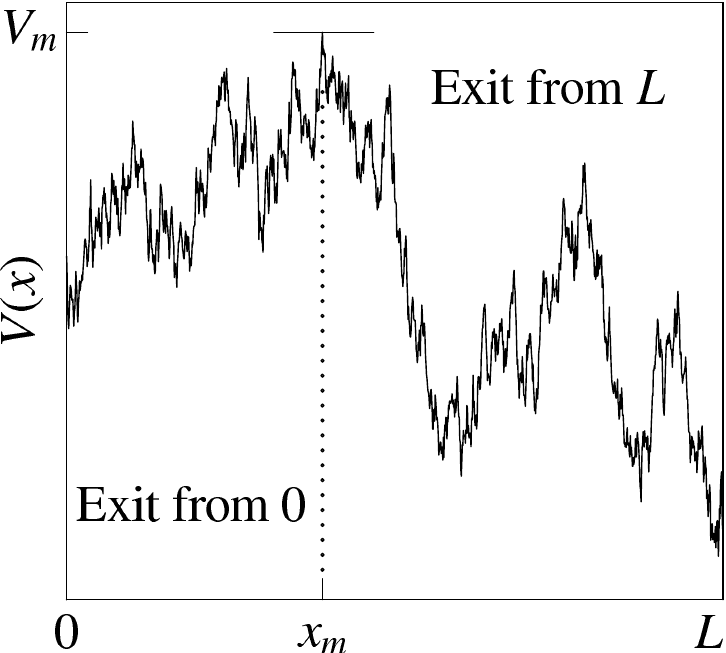}}
\caption{Left. A stochastic process starting at $x$ 
leaves the positive half axis for the first time at $t_f$; 
$m$ denotes its maximum till $t_f$. Right. 
A self-affine disordered potential with maximum at $x_m$: 
when $L$ is large, the diffusing particle, starting at $0<x<L$, exits 
the box through $0$ for $x<x_m$ and through $L$ for $x>x_m$.}
\label{fig:max1}
\end{figure}

{\em Self-affine processes.} To compute $Q(x,L)$ in a box geometry, 
it is useful 
first to relate it to another quantity associated with the 
same process $X(t)$, but 
now in a semi-infinite geometry $[0,\infty]$. Consider a process $X(t)$ 
in $[0,\infty]$, starting at $x$ and absorbed at the origin for the
first time at $t_f -$ the first-passage time. Let $m$ 
denote the maximum of this process till $t_f$ 
(see Fig.~\ref{fig:max1} left). Then, 
it is clear that $1-Q(x,L)$, the probability that the particle 
exits the box through 
the origin (and not through $L$), is precisely equal to the 
probability that the maximum $m$ of the process in $[0,\infty]$
till $t_f$ stays below $L$, i.e.,
the cumulative 
distribution of 
$m$, ${\rm Prob}[m\le L|x]$, in the semi-infinite geometry. 
The distribution of $m$ is, in 
turn, related to the distribution of the first-passage time $t_f$. 
Let $q(x,T)={\rm 
Prob}[t_f \ge T|x]$ denote the cumulative probability of 
$t_f$, which is also the 
survival probability of the particle starting at $x$ in the 
semi-infinite geometry. 
One knows that for generic self-affine processes 
$q(x,T)= q(x/T^H)$. For large $T$, 
$q(x,T)\sim T^{-\theta}$, where $\theta$ is the persistence exponent of the 
process~\cite{pers_review}. This implies the scaling function $q(y)\sim 
y^{\theta/H}$ for small $y$~\cite{ZRM}. 
Noting that $m\sim t_f^H$ for self-affine 
processes, it follows that 
$Q(x,L)=1-{\rm Prob}[m\le L|x]= {\rm Prob}[m\ge L|x]\approx 
{\rm Prob}[t_f\ge L^{1/H}|x]=q[x/L]$. This demonstrates the scaling 
behavior anticipated 
before, namely, $Q(x,L)=Q(x/L)$, where $Q(z)=q(z)$. Moreover, since $q(y)\sim 
y^{\theta/H}$ for small $y$, we get $Q(z)\sim z^{\phi}$ for small $z$, with 
$\phi=\theta/H$. For example, for Brownian motion $H=1/2$ and $\theta=1/2$, hence 
$\phi=1$, in accordance with the exact result $Q(z)=z$. For the subclass of 
self-affine processes with stationary increments, the same exponent 
$\phi$ happens 
to describe the vanishing of the probability density close to an absorbing 
boundary~\cite{ZRM}.

\begin{figure}
\centerline{
\includegraphics[width=\columnwidth]{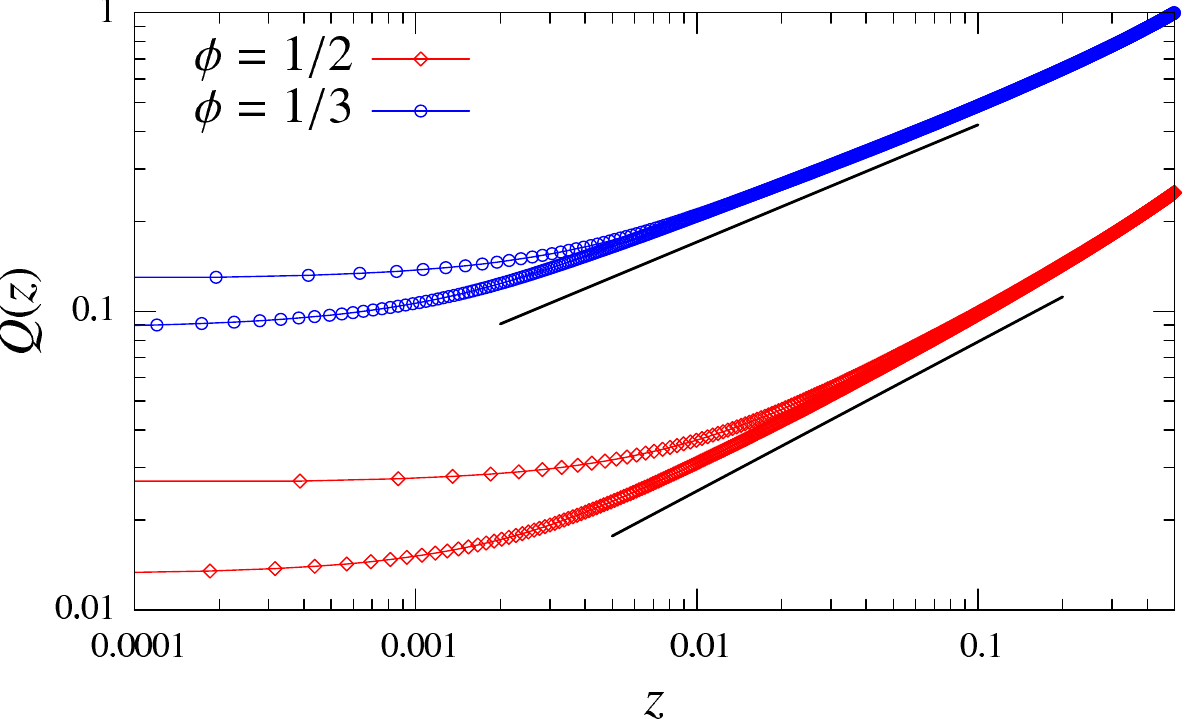}}
\caption{Behavior of $Q(z)$ close to $z=0$ for fBm processes. For $H=2/3$ ($\phi = 1/2$) the size of the box is $L=50$, $200$; for $H=3/4$ ($\phi = 1/3$) the size of the box is $L=100$, $300$. The continuum limit is reached when $L\rightarrow \infty$. The expected slopes are reported as solid lines. Data have been shifted to make visualization easier.}
\label{fig:exponent}
\end{figure}

Our general prediction $Q(z)\sim z^\phi$ for small $z$ can
be verified explicitly for some self-affine processes where
$Q(z)$ can be computed exactly, as discussed later.
Moreover, we have numerically verified that this conjecture holds 
also for the fractional Brownian motion (fBm), i.e., a self-affine 
Gaussian process defined by the following autocorrelation function
\begin{equation}
\langle X(t_1) X(t_2)\rangle = \frac{1}{2} \left( t_1^{2 H}+ t_2^{2H} - |t_1-t_2|^{2 H}\right),
\end{equation}
with $0 <H < 1$~\cite{mandelbrot, pers_interface}. 
In~\cite{ZRM}, we have proposed fBm as a natural candidate for describing
the time evolution of the  
translocation coordinate. For this process, the persistence exponent is 
known, $\theta=1-H$ \cite{pers_interface}, so that $\phi=(1-H)/H$. An 
expedient algorithm for generating fBm paths is provided in~\cite{dieker}. 
The probability $Q(z)$ can be numerically computed as follows. Given a 
realization of the process starting from the origin, we record its minimum and 
maximum values for increasing time; the process is halted when 
$X_\text{max}-X_\text{min} \ge L$. If the last updated quantity is 
$X_\text{min}$, the contribution to $Q(x,L)$ is $0$ for $x \in (0, L-X_\text{max})$ 
and $1$ for $x \in (L-X_\text{max}, L)$. In the opposite case, the contribution 
is $0$ for $x \in (0, -X_\text{min})$ and $1$ othewise. 
All simulations are performed by averaging over $10^6$ samples. 
Fig.~\ref{fig:exponent} shows the agreement between numerical simulations 
and predicted scaling of $Q(z)$ for different values of the parameter 
$H$.\\

{\em Disordered potential.} We next consider the 
stochastic motion of a single particle diffusing in a potential $V(X)$,
starting at the initial position $x$. 
The dynamics is governed by the Langevin equation $\dot{X}(t) = f[X(t)] + \eta(t)$, where $X(0)=x$ and $f(X) = -dV (X)/dX$ is the force and $\eta(t)$ is a 
Gaussian white noise 
with $\langle \eta(t)\rangle =0$ and 
$\langle \eta(t)\eta(t')\rangle = \delta(t-t')$. 
To compute $Q(x,L)$, the first step is to 
write a differential equation satisfied by $Q(x,L)$, 
taking the initial position $x$ as a variable and keeping the box size $L$ 
fixed. 
Let us consider a small time interval $[0,\Delta t]$ at the 
beginning of the process. In this time interval, the particle moves 
from its initial position $x$ to a new position 
$x + \Delta x$ at time $\Delta t$, 
where $\Delta x = f(x)\Delta t + \eta (0) \Delta t$, 
$\eta(0)$ being the noise variable that kicks in at time $0$. 
Since the process is Markovian, the subsequent evolution does not 
know about the interval $[0,\Delta t]$, hence one gets
\begin{equation}
Q(x,L) = \langle Q(x + f(x)\Delta t + \eta(0) \Delta t,L) \rangle,
\label{qmx}
\end{equation}
where $\langle \rangle$ denotes the average over the 
initial noise $\eta(0)$. Expanding the rhs of 
Eq.~\eqref{qmx} as a Taylor series in powers of 
$\Delta t$, using $\langle \eta (0)\rangle = 0$ and 
$\langle \eta^2(0)\rangle = 1/\Delta t$ (delta correlated noise),
yields an ordinary differential equation, $\frac{1}{2} Q''(x)+f(x) Q'(x)=0$.
Solving with 
boundary conditions $Q(0,L)=0$ and $Q(L,L)=1$ gives the exact result
\begin{equation}
Q(x,L) = \frac{\int_0^x e^{2V(x')}dx'}{\int_0^L e^{2V(x')}dx'},
\label{qmx_V}
\end{equation}
valid for arbitrary potential $V(X)$. 
Note that for a potential-free particle, i.e., 
$V(X) = 0$ in Eq.~\eqref{qmx_V}, we recover the Brownian 
result, $Q(x,L) = x/L$.

\begin{figure}
\centerline{
\includegraphics[width=\columnwidth]{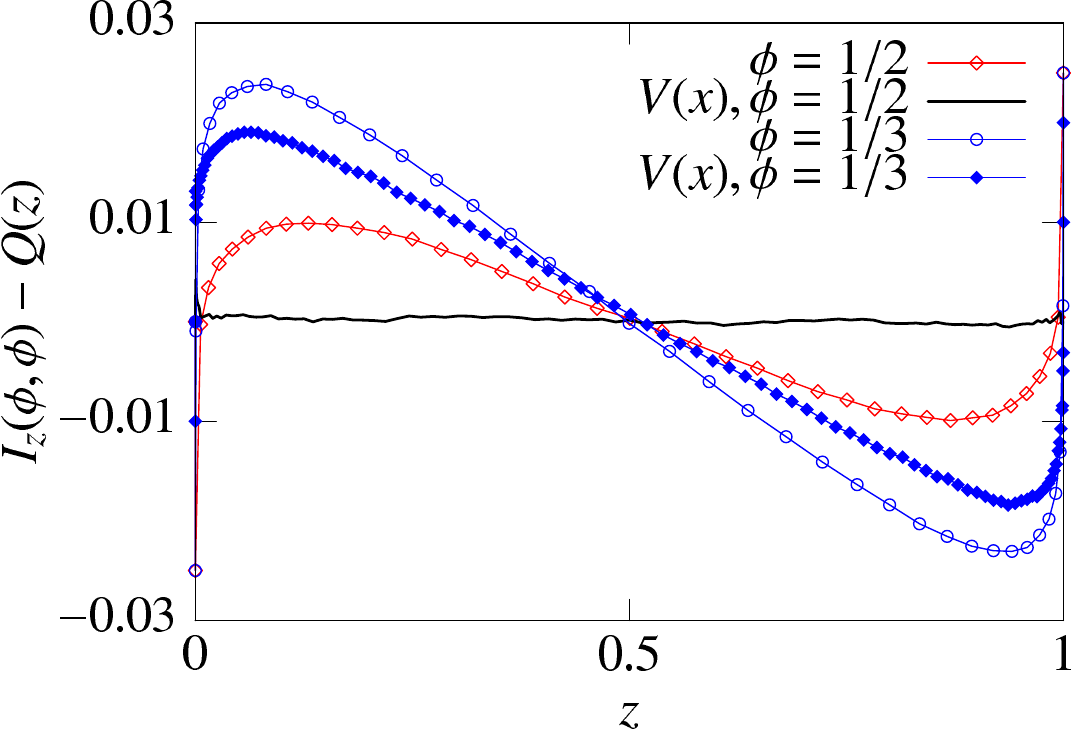}}
\caption{The difference between Eq.~\eqref{incomplete_beta} and simulated $Q(z)$. For fBm processes: $H=2/3$ ($\phi=1/2$) with box size $L=200$, and $H=3/4$ ($\phi=1/3$), with box size $L=300$. For fBm disordered potentials: $H_V=2/3$ ($\phi=1/3$), with box size $L=10^4$. For comparison, we display also the Sinai model $H_V=1/2$ ($\phi=1/2$), with box size $L=10^4$.}
\label{fig:difference}
\end{figure}

Taking derivative with respect to $x$ gives 
\begin{equation} 
p_\text{eq}(x,L)=\frac{\partial}{\partial x}Q(x,L) = 
\frac{e^{2V(x)}}{\int_0^L e^{2V(x')}dx'}, 
\label{peak} 
\end{equation} 
which can be interpreted as the equilibrium probability density of the particle 
to be at $x$ in presence of a potential $-V(X)$. When $V(X)$ is a 
realization of a 
disordered potential, it is natural to introduce $\overline{Q(x,L)}$, the 
disorder-averaged hitting probability. An example where we can determine 
$\overline{Q(x,L)}$ explicitly is the classical Sinai model, i.e., when the 
potential $V(X)$ is a trajectory of a Brownian motion in space, $V(X) \sim 
X^{1/2}$~\cite{sinai}. For this model the $\overline{p_\text{eq}(x,L)}$ can 
be computed exactly~\cite{texier, broderix} 
\begin{equation} 
\overline{p_\text{eq}(x,L)}=\frac{1}{\pi}\frac{1}{\sqrt{x(L-x)}}. 
\label{p_max} 
\end{equation} 
Thus, $\overline{Q(x,L)}=Q(z=x/L)$ again 
satisfies the generic scaling with a scaling function form with 
\begin{equation} 
Q(z) = \frac{2}{\pi} \arcsin \left( \sqrt{z} \right). 
\label{sinai} 
\end{equation} 
Note that close to the origin $Q(z) \sim 
z^\phi$ with $\phi=1/2$. On the other hand, it is well known that in the 
Sinai potential the particle evolves very slowly with time, $X \sim 
\ln^2(t)$, showing a self-affine scaling in the variable $T=\log t$, with 
$H=2$. For this model, it is also known that the survival probability 
decays as $1/\log t$, i.e., $T^{-\theta}$, with $\theta=1$\cite{comtet, 
pierre}. Thus, $\theta/H=1/2=\phi$, in accordance with our general scaling 
prediction.

We next consider a generic self-affine potential, $V(X)\sim X^{H_V}$ 
(with $V(0)=0$), the Sinai potential being a 
special case with $H_V=1/2$.
We show that $\overline{p_\text{eq}(x,L)}$ for such a potential is related
to the probability density of the location $x_m$ of the maximum
of the potential $V(X)$ over $X\in [0,L]$.
We rewrite Eq. (\ref{peak}) as $p_\text{eq}(x,L)=[\int_0^L 
e^{2(V(x')-V(x))}dx']^{-1}$, rescale variables $x'\to x'L$ and $x\to xL$
and use the self-affine property $V(xL)=L^{H_V}V(x)$ to obtain
$p_\text{eq}(x,L)=[\int_0^1 e^{2L^{H_V}(V(x')-V(x))}dx']^{-1}$.
For large $L$, using a steepest decent method, we immediately
see that, for each realization of the disorder potential $V(X)$,
$p_\text{eq}(x,L)\simeq \delta(x-x_m)$ where $x_m$
denotes the position where $V(X)$ is maximum.
This observation has two immediate consequences: (i) By integrating over $x$,
we get, for each realization, $Q(x,L)\simeq \theta(x-x_m)$. This means
that, for any given realization, if the starting position $x$ is to the left 
(right) of the location
$x_m$ of the maximum, $Q(x,L)\simeq 0$ (respectively $Q(x,L)\to 1$) indicating 
that the
particle exits the box through $0$ (through $L$) as 
depicted in Fig.~\ref{fig:max1}(right). (ii) By taking average over the 
disorder, we get   
\begin{equation}
\overline{p_\text{eq}(x,L)}\simeq p_m(x,L)
\label{pmax1}
\end{equation}
where $ p_m(x,L)$ is the probability density that the maximum of the potential 
$V(X)$ over $[0,L]$ is located at $x$.
For example, for the Sinai
case, one knows from L\'evy's
arcsine law~\cite{Feller} that $p_m(x,L)= 1/{\pi \sqrt{x(L-x)}}$.
Thus, in this case the relation (\ref{pmax1}) is 
verified by the exact result (\ref{p_max}). However, the
relation (\ref{pmax1}) is more general and holds for arbitrary
self-affine potential. We remark here that physically the relation 
(\ref{pmax1}) reflects the fact that in a self-affine potential
$-V(X)$ in which the particle is at equilirium, 
the limit $L\to \infty$ limit is equivalent to the zero temperature
$T\to 0$ limit forcing the particle to the minimuum of the potential $-V(x)$
or equivalently to the maximum $x_m$ of $V(x)$.

A useful consequence of (\ref{pmax1}) is that it allows us
to relate the persistence or the survival probability of a particle 
moving in a disordered self-affine potential to the statistical
properties of the potential $V(X)$ itself. 
The disordered potential 
$V(X) \sim X^{H_V}$ (we assume $V(0) = 0$)
can itself be regarded as a stochastic process with the space coordinate $X$ 
playing the role 
of `time'. So, the probability that $V(X)$ stays below (or above) 
the
level $X=0$ up to a distance $L$ decays, for large $L$, as $L^{-\theta_V}$
where $\theta_V$ is the spatial persistence exponent~\cite{spatpers} of 
$V(X)$.
For example, for the Sinai potential (brownian motion in space), 
$\theta_V=1/2$. The pair of exponents $(H_V,\theta_V)$ associated
with the potential can now be related to the corresponding
exponents associated with the temporal motion of the particle
in this potential.   
By Arrhenius' law for the activated dynamics, the time required 
for 
particle diffusing in $V(X)$ to overcome an energy barrier scales as $t \sim 
e^{V(X)}$. Using $V(X)\sim X^{H_V}$, one deduces that $X \sim 
T^{1/H_V}$ where $T=\log(t)$. Thus the particle motion $X(T)\sim T^{H}$ is a 
self-affine process as a function of $T=\log(t)$, with a Hurst 
exponent $H=1/H_V$.
Next, we note that $p_m(x,L)$, the probability that the maximum
of $V(X)$ occurs at $x$, coincides, when $x\to 0$, with 
the probability that $V(X)$ stays below $0$ up to a distance $L$,
hence $p_m(x\to 0, L)\propto L^{-\theta_V}$. On the other hand,
based on our general argument, we expect that ${\overline {Q(x,L)}}\sim 
(x/L)^{\phi}$ when $x\to 0$, where $\phi=\theta/H$. 
this means that $\overline{p_\text{eq}(x,L)}\propto x^{\phi-1}/L^{\phi}$.
Note
that here $\theta$ is the persistence
exponent associated with the temporal motion of the particle, i.e.,
the survival probability of the particle up to time $T=\log(t)$ decays
as $\sim T^{-\theta}$.
Matching powers of $L$ from both sides of (\ref{pmax1}) provides
the desired relation between temporal and spatial exponents
$\theta=\theta_V H= \theta_V/H_V$. 
For instance, in the 
Sinai model, using $\theta_V=1/2$, $H_V=1/2$ we get and $\theta = 1$, in
agreement with the exact result~\cite{comtet,pierre}. 
If one considers a potential $V''(X)=\xi(X)$ where
$\xi(X)$ is a white noise in space, it is self-affine
with $H_V=3/2$. The exponent $\theta_V=1/4$ is known exactly~\cite{burkhardt}.
Thus we predict that for this poential, the survival probabaility
up to time $t$ will decay as $\sim (\log t)^{-\theta}$ with
$\theta=1/6$.

{\em Super-universality of $Q(z)$.} For some non-Brownian stochastic 
self-affine processes, 
the full function $Q(z)$ is known. 
For instance, L\'evy Flights are Markovian superdiffusive processes 
whose increments obey a L\'evy stable (symmetric) law of 
index $0<\mu \le 2$. The Hurst exponent is $H=1/\mu$. By virtue of the 
Sparre Andersen theorem~\cite{sparre}, the persistence exponent is 
$\theta=1/2$, independent of $\mu$. Hence, $\phi=\theta/H=\mu/2$ 
(see also~\cite{zumofen,zoia_rosso}). The full function $Q(z)$
for L\'evy Flights has been 
computed~\cite{BGR} and can be recast in an elegant form 
\begin{equation}
Q(z)=I_z(\phi,\phi)= \frac{\Gamma(2\phi)}{\Gamma^2(\phi)} 
\int_{0}^{z} \left[ u(1-u) \right]^{\phi-1} du,
\label{incomplete_beta}
\end{equation}
i.e., a regularized incomplete Beta function containing a single parameter $\phi=\mu/2$. Clearly, $Q(z)\sim z^{\phi}$ as $z\to 0$ in agreement with our prediction. The formulae for Brownian motion (with $\phi=1$) $Q(z)=z$ and for the Sinai model ($\phi=1/2$) in \eqref{sinai} can also be expressed as~\eqref{incomplete_beta}. Moreover, the distribution of the maxima for a symmetric L\'evy Flight process is given by~\eqref{p_max}, by virtue of the Sparre Andersen theorem~\cite{sparre}. Hence, we expect the hitting probability~\eqref{incomplete_beta} to apply also to particles diffusing in a L\'evy Flight disordered potential, with $\phi=1/2$. Finally, $Q(z)$ is known also for the Random Acceleration model, a non-Markovian process that is defined by $d^2 X / dt^2= \eta(t)$, with $\eta(t)$ as before. The motion starts at $X(0)=x$, with initial velocity $v(0)=0$, and is superdiffusive, with $X \sim t^{3/2}$, i.e., $H=3/2$. Its first-passage properties have been widely studied~\cite{burkhardt}. The persistence exponent is $\theta=1/4$, so that $\phi=\theta/H=1/6$. Bicout and Burkhardt~\cite{BB} also computed the full exit probability $Q(x,L)$. One can again recast this formula in the same super-universal form \eqref{incomplete_beta} with $\phi=1/6$.

Based on these special cases, it may be tempting to conjecture that the full function $Q(z)$ for arbitrary anomalous doffusion processes has the super-universal form ~\eqref{incomplete_beta}, the only information about the process enters this formula through a single exponent $\phi$. However, this turns out not to be the case, and we are able to show notable counterexamples. In Fig.~\ref{fig:difference}, we compute the hitting probability for fBm self-affine processes and for particles diffusing in fBm disordered potentials, and display the numerical difference with respect to formula~\eqref{incomplete_beta}, with the appropriate exponent $\phi$. We find that in neither case $Q(z)$ can be described by the super-universal form~\eqref{incomplete_beta}.

\end{document}